# Prisoner's Dilemma in *Maximization constrained: the rationality of cooperation*

*Shahin Esmaeili*

## I

 David Gauthier in his article, *Maximization constrained: the rationality of cooperation*, tries to defend of the joint strategy in situations which no outcome is both equilibrium and optimal. Prisoner's Dilemma is the most familiar example of these situations. He first starts with some quotes by Hobbes in *Leviathan*; Hobbes, in chapter 15 discusses an objection by someone is called Foole, and then will reject his view. In response to Foole, Hobbes presents two strategies (i.e. joint and individual) and two kinds of agents in such problems including Prisoner's Dilemma – i.e. straightforward maximizer (SM) and constrained maximizer(CM). Then he considers two arguments respectively for SM and CM, and he will show that why in an ideal and transparent situation, the first argument fails and the second one would be the only valid argument. likewise, in the following part of his article, he considers more realistic situations with translucency and he concludes that under some conditions, joint strategy would be still the rational decision.

## 1.1 Hobbes and Foole

In chapter 15 of *Leviathan*, Hobbes states:

> "The Foole hath sayd in his heart, there is no such thing as Justice; and sometimes also with his tongue; seriously alleging, that every mans conservation, and contentment, being committed to his own care, there could be no reason, why every man might not do what he thought conduced thereunto: and therefore also to make, or not make; keep, or not keep Covenants, was not against Reason, when it conduced to ones benefit." (Gauthier, 1990, P 315)

Foole believes that a cooperative strategy is acceptable only if you "expect the agreement to pay", and he agrees to adhere a cooperative one only if it increases the utility and proves the maximized expected utility. Foole states:

> "Foole insists that for it to be rational to comply with an agreement to cooperate, the utility an individual may expect from cooperation must also be no less than what he would expect were he to violate his agreement. [no less than ideal] And he then argues that for it to be rational to agree to cooperate, then, although one need not consider it rational to comply oneself, one must believe it rational for the others to comply." (Gauthier, 1990, P 316 )

And according to Gauthier, Fool's argument is as follows:

> "That one had reason for making an agreement can give one reason for keeping it only by affecting the utility of compliance. To think otherwise is to reject utility-maximization. He insists that holders of this view have failed to think out the full implications of the maximizing conception of practical rationality." (Gauthier, 1990, P 318)

But in contrary with Foole, Hobbes believes that it is not reasonable to be unjust, because it is not reasonable to believe that doing so you can maximize your utility. [1]

Gauthier refutes Foole's idea and believes that if it is rational to make an agreement, it would be rational to keep it. He first induces two main strategies, and Using Modus Tollens, he shows how Foole's approach is not a real cooperation.

Gauthier starts his argument with a distinction between individual and joint strategy. He states that and an "individual strategy is a lottery over the possible actions of a single actor." And a joint strategy is "a lottery over possible outcomes." What he means by 'possible outcomes' is all possible options for all cooperators in decision situation. To shed light on joint strategy, he gives one example about hunting in which each hunter has some particular responsibility coordinating with those of others leading to a favorable result.

Gauthier stresses that a person decides to cooperate with his fellows only if he bases his actions on joint strategy and such person would be committed to all implications of his decision. Hobbes' argument can be formulated as follows:

- If a person agrees to cooperate with his fellows, then it means that he will employ the joint strategy and he will keep it.
- But the person (who makes an agreement) doesn't comply himself to keep it,
- Then, it means that he doesn't agree to cooperate with his fellows.

  So, Foole doesn't really believe to cooperation, and he act just base on his individual strategy.

After showing that Foole's idea is really just acting based on individual strategy, Hobbes argues that individual strategy is not the rational decision.

He starts with introducing two kinds of agents (based on two kinds of strategies) and he will consider different arguments on Prisoner's Dilemma. (Gauthier 1969)[2]

## 1.2 Straightforward maximizer (SM) and constrained maximizer (CM)

---

[1]. For his view on moral reason see, (Gauthier 1986)

[2]. For a historical debate on Prisoner's dilemma see: (Carroll, 1987; Axelrod 1988; Aumann 1995; Farrell 1989; Howard 1988; Kraines 1989; Lewis 1979; Poundstone1992; Kreps 1982; Kendall 2007; Batali 1995; Bendor 1987)

According to Gauthier, a straightforward maximizer (SM) "is a person who seeks to maximize his utility given the strategies of those with whom he interacts." And on the other hand, a constrained (CM) "is a person who seeks in some situations to maximize her utility, given not the strategies but the utilities of those with whom she interacts." 319 It means that SM thinks of her partners' strategies so that she can exploit them using the best strategy against them. He thinks about the strategies just to make a chance to bitrate her partners. In contrary, a CM just concentrates on how they (i.e. herself and her partners) can maximize their utilities using best joint strategy. (Afroogh 2021)

So, Hobbes and Gauthier both accept the rationality of joint strategy, and Foole believes the rationality of individual maximization. However, Gauthier doesn't believe to an Absolut cooperation or joint strategy. He defends of a constrained maximizer who bases her action on joint strategy just under some conditions.

A constrained maximizer is an agent who conditionally disposed to act. Gauthier defines it as an agent who acts based on joint strategy except two situations:

> 1- "a constrained maximizer does not base her actions on a joint strategy whenever a nearly fair and optimal outcome would result were everyone to do likewise." [everyone does straightforwardly]
> 2- "Faced with persons whom she believes to be straightforward maximizers, a constrained maximizer does not play into their hands by basing her actions on the joint strategy she would like everyone to accept, but rather, to avoid being exploited, she behaves as a straightforward maximizer, acting on the individual strategy that maximizes her utility given the strategies she expects the others to employ." (Gauthier, 1990, p 319)

> Moreover, Gauthier adds that "A constrained maximizer is conditionally disposed to act, not only on the unique joint strategy that would be prescribed by a rational bargain, but on any joint strategy that affords her a utility approaching what she would expect from fully rational co-operation. (approaching the outcome determined by minimax relative concession)" (Gauthier, 1990, p 319 )

II

## 2.1 The familiar Prisoner's Dilemma: who is right?

In the familiar Prisoner's Dilemma, a SM like Foole chooses not to cooperate and a CM like Gauthier believes to cooperate with his partner. But who is right? each of them think that her strategy is the best way to maximize her expected utility. So, it seems that the best way to evaluating their claims is to calculate the expected utility of each strategy. Gauthier choose this solution and present two arguments based on calculating the expected utility of individual and joint strategy.

## 2.2Two arguments: The expected utility of SM and CM

Gauthier supposes utility of *u* for each person act on an individual strategy, and utility of *u'* for a cooperative joint strategy, and utility of *u"* for an individual strategy who her partners base their actions on a cooperative joint strategy. So we can show Expected utilities of three situations as follows:

U < U' < U"
SM<CM<one SM and others are CMs

Moreover, he assumes that P refers to the probability that others are CMs. We can summarize his first argument for SM as follows:

1- Suppose I adopt straightforward maximization. Then,
 If I expect the others to cat as CM ---> I get u"
 If I expect the others to act as SM ---> I get u
Therefore, my overall expected utility is [pu" + (1-p)u].
2- Suppose I adopt constrained maximization. Then,
If I expect the others to cat as CM ---> I get u'
 If I expect the others to act as SM ---> I get u
Therefore, my overall expected utility is [pu' + (1-p)u].
3-Since u" is greater than u',[pu" + (1-p)u] is greater than [pu' + (1-p)u], Therefore, to maximize my overall expectation of
utility, I should adopt straightforward maximization. (Gauthier, 1990, p 322 )

We can show the expected utility of SMs in matrix (1).

**Matrix (1) for Argument (1) or dominance argument for SM**

|  |  |  | Prisoner 2 |  | My overall EU |
|---|---|---|---|---|---|
|  |  | CM | SM |  |  |
| Prisoner 1 (me) | SM | I get U" | I get U |  | pu" + (1-p)u |
|  | CM | I get U' | I get U |  | pu' + (1-p)u |

But this argument is not sound. Because it presupposes that no person knows other's disposition. In this argument this is assumed that you can be SM at the same time that you partners are CMs, and they are not aware of your dispositions. However, by definitions, CM is a person who conditionally disposed to act, and as we mentioned, she bases her act based on joint strategy only with those whom they supposed to be CMs. So, a SM can't exploit a CM in this case. Therefore, this argument for SM fails.

Likewise, assuming the followings keys, we can calculate expected utility of CMs:
 U < U' < U"
SM<CM<one SM with all CMs
Moreover, he assumes that P refers to the probability that others are CMs.
We can formulate his argument for CM as follows:

1- Suppose I adopt straightforward maximization. Then,
I must expect the others to employ SM --- > my expected utility is *u.*
2- Suppose I adopt constrained maximization. Then,
If the others are conditionally disposed to CM ---> I may expect them to base their actions on a co-operative joint strategy, and my expected utility is *u'.*
If they are not so disposed [are not CM], ---> I employ SM and expect *u* as before.
Therefore, If the probability that others are disposed to constrained maximization is *p,* then my overall expected utility is *[pu' + (1-p)u].*
3- Since *u'* is greater than *u,[pu' + (1-p)u]* is greater than *u.* Therefore, to maximize my overall expectation of utility, I should adopt constrained maximization. (Gauthier, 1990, p 323)

**Matrix (2) for Argument (2) for CM.**
X stands for "prisoner 2 dosent act based on the relevant strategy"

|  |  |  | Prisoner 2 |  | My overall EU |
|---|---|---|---|---|---|
|  |  | SM | CM |  |  |
| Prisoner 1 (me) | SM | I get U | X |  | U |
|  | CM | I switch to SM state and get U | I get U' |  | *pu' + (1-p)u* |

Argument (2) takes into account what argument (1) ignores. Gautier states that argument (2) takes into account what argument (1) ignores. Constrained maximizer interacts cooperatively just with CMs, and they employ individual strategy if they faced with SMs. So, "constrained maximizers are able to make beneficial agreements with their fellows that the straightforward cannot, not because the latter would be unwilling to agree, but because they would not be admitted as parties to agreement given their disposition to violation.
Straightforward maximizers are disposed to take advantage of their fellows should the opportunity arise; knowing this, their fellows would prevent such opportunity arising." (Gauthier, 1990, p 323)
However, SM can benefit by using a guise with the appearance of constrained maximizer, and they can exploit CMs by deception. Gauthier answers this objection emphasizing that joint strategy are applied to ideal persons and he adds that we can propose another idealizing constraint and say that all these persons are transparent and because of this any deception is impossible.
It seems that the objection is answered, however, a more important objection gives raise. If we assume such an idealized argument, we can't apply it on decision situations in real world. In other words, if we want to have more realistic decision we should present a less ideal argument with less special constraints.

## 2.3 The expected utility of SM and CM in translucent cases

That said, Gauthier begins to argue for less transparent cases which he call as translucent ones and he tries to calculate the expected utility of CMs and SMs in these cases. He defines translucency as *"supposing that persons are neither transparent nor opaque, so that their disposition to co-operate or not may be ascertained by others, not with certainty, but as more than mere guesswork."* (Gauthier, 1990, p 324)

For answering the question that "under which conditions it would be rational to dispose oneself to constrained maximization?" Gauthier defines four possible situations as follows:

1- Non-cooperation: [SM] if it is not the case of cooperation and exploitation. (value= U')

 2- cooperation :those interacting are CMs who achieve mutual recognition, in which case the co-operative outcome results (value= U")

3- Defection: we ignore (this case) the inadvertent taking of advantage when CMs mistake their fellows for SMs. (value= 1)

4- Exploitation: those interacting include CMs who fail to recognize SMs but are themselves recognized, in which case the outcome affords the SMs the benefits of individual defection and the CMs the costs of having advantage taken of mistakenly basing their actions on a cooperative strategy. (value= 0)

We can present these utility in the following sequence:
 Defection (1)> Cooperation(U")>Non-cooperation (U')>Exploitation (0)

Three probabilities:
P: Probability that CMs will achieve mutual recognition and so successfully cooperate.
q: Is the probability that CMs will fail to recognize SMs but will themselves be recognized, so that defection and exploitation will result.
R: Is the probability that a randomly selected member of the population is a CM. (so the probability that a randomly selected person is an SM is *(1-r)*.
The values of *p,q,* and *r* must of course fall between 0 and 1.

**CM's expected utility in translucent case**

|  |  |  | Prisoner 2 |  |  |  |
|---|---|---|---|---|---|---|
|  |  | SM or CM who are not recognized | CM | SM |  |  |
| Prisoner 1 (me) | CM | I do individually | Both succeed in cooperating | I am exploited |  |  |
| My EU |  | +U' | (u"-u') over U' | -[(1-r)qu'] | = | *{u'+ [rp(u"-u')] - (1-r)qu'}.* |

A CM expects the utility *u' [SM]* unless (1) she succeeds in co-operating with other CMs [*u"*] or (2) she is exploited by an SM. [0]
Probability of (1): rp= (probability that she interacts with a CM). (probability that they achieve mutual recognition)

Value of (1): *(u″-u′)* over her non-co-operative expectation *u′*
The effect of (1) = is to increase her utility expectation by a value *[rp(u″-u′)]*
The probability of (2): *(1-r)q =* (probability that she interacts with a SM ) . *(the probability* that she fails to recognize him [SM] but is recognized*)*
Value of (2)=0
The effect of (2): is to reduce her utility expectation by a value *[(1-r)qu′]*
CM expected utility = *Utility u′ (SM) + effect of CM+ effect of being Exploited= {u′+ [rp(u″-u′)] - (1-r)qu′}.*
(Gauthier, 1990, p 325)

**SM's expected utility in translucent case**

| | | | Prisoner 2 | | |
|---|---|---|---|---|---|
| | | SM | CM | | |
| Prisoner 1 (me) | SM | Bothe do individually | I exploit her | | |
| Value | | U' | *(1-u′) over U'* | | |
| My EU | | +U' | +[rq(1-u')] | = | *{u′ + [rq(1-u′)]}* |

An SM expects the utility u' unless he exploits a CM.
The probability of exploitation: (the probability that he interacts with a CM) . (hat he recognizes her but is not recognized by her)= rq
Value of exploitation: *(1-u′)* over his non-co-operative expectation *u′*.
The effect of exploitation: is to increase his utility expectation by a value *[rq(1-u′)]*
SM's expected utility = utility of SM+ the effect of exploitation= *{u′ + [rq(1-u′)]}* (Gauthier, 1990, p 326)

**When constrained maximization is rational in translucent cases?**
As it clear, the result of comparison in translucent cases is not straightforward and we cannot say generally which strategy is always rational.

And we can say that "It is rational to dispose oneself to constrained maximization if and only if the utility expected by a CM is greater than the utility expected by an SM, which obtains if and only if *p/q* is *greater* than *[(1-u′)/(u″-u′) + [(1-r)u′]/[r(u″-u′)]}.''* (Gauthier, 1990, p 326)
Or we can say that acting based on cooperation is rational only if "the ratio of p to q, i.e. the ratio between the probability that an interaction involving CMs will result in co-operation and the probability that an interaction involving CMs and SMs will involve exploitation and defection, is greater than the ratio between the gain from defection and the gain through co-operation." (Gauthier, 1990, p 326)

**2.4. Objection to our argument and answer**

That said, we can conclude that the expected utility of CMs would be greater than SMs generally in transparent case and partially in translucent ceases. However, it doesn't explicitly mean that joint strategy would be rational choice. For to be rational choice the expected utility should be nearly optimal and fair one. Gauthier response and states that our argument *implicitly* means that expected utility of CMs are nearly optimal and fair one. Moreover, he presents a new distinction and adds a new constraint. He states that there two kinds of disposition of CMs:

>*Narrowly compliant:" a* person who is disposed to cooperate in ways that, followed by all, yield nearly optimal and fair outcomes*."*
>
>*Broadly compliant:"* a person who is disposed to co-operate in ways that, followed by all, merely yield her some benefit in relation to universal non-cooperation" (Gauthier, 1990, p 328)

He argues that in many situations a broadly compliant would lose. Because in so far as she is known as broadly compliant, other persons will exploit her and they will maximize her utilities by offering a cooperative action with just a little benefit for the broadly compliant person. Goutier add a new constraint for CMs and states we mean by CMs just narrowly compliant persons.

In the last part of his article, Gautier present a supplementation for his arguments and states that if we want to benefit from joint strategy, it is necessary for us to develop our ability of detecting other's dispositions. He stresses that "failure to develop this ability, or neglect of its exercise, will preclude one from benefiting from constrained maximization. And it can then appear that constraint is irrational. But what is actually irrational is the failure to cultivate or exercise the ability to detect others' sincerity or insincerity." (Gauthier, 1990, p 329)

## III

## Conclusion

Gauthier believes that constrained maximization is the rational strategy based on calculating EUs concentrating on the different strategies which are taken by decision maker and the other agents. He didn't pay attention to the difference of outcomes and entries and how we evaluate them by CEU and EEU. He sees the CM as a constant strategy which works in all cases (generally in transparent cases and partially in translucent cases) no matter what the outcomes are, what kinds of relationships are there, and what the context is. Gauthier's rational theory in Prisoner's Dilemma works just in special cases including some ideal agents, or in less ideal cases but in an isolated context – I mean the contexts which ignore any relationship between actions and outcomes. In contrary, Nozick denies cooperation (1993 & 1969). For a contextualist approach in rationality see (Afroogh 2019; 2020).